\documentstyle[twocolumn,prl,aps,epsf,floats]{revtex} 
\begin{document} 
\title{Many-body Effects in Angle-resolved Photoemission: Quasiparticle Energy 
and Lifetime of a Mo(110) Surface State} 
\author{T. Valla, A. V. Fedorov and P. D. Johnson}
\address{Department of Physics, Brookhaven National Laboratory, Upton, NY, 
11973-5000}
\author{S. L. Hulbert}
\address{National Synchrotron Light Source, Brookhaven National Laboratory, 
Upton, NY, 11973-5000}
\address{ {\em \bigskip \begin{quote}
In a high-resolution photoemission study of a Mo(110) surface state various contributions to the measured width and energy of the quasiparticle peak are 
investigated. Electron-phonon coupling, electron-electron interactions and 
scattering from defects are all identified mechanisms responsible for the finite 
lifetime of a valence photo-hole. The electron-phonon induced mass enhancement 
and rapid change of the photo-hole lifetime near the Fermi level are observed 
for the first time.
\end{quote}}}
\maketitle
Recent investigations of strongly correlated electron systems have questioned 
the validity of one of the most fundamental paradigms in solid state physics - 
Fermi liquid theory. The latter picture is based on the existence of 
"quasiparticles", or single-particle-like low energy excitations which obey the 
exclusion principle and have lifetimes long enough to be considered as 
particles. Strictly speaking, the quasiparticle concept is restricted to zero 
temperature and a narrow region around the Fermi level \cite{1}, but its 
usefulness often continues to finite temperatures, and energies away from the 
Fermi level \cite{2}. Indications for possible non-Fermi-liquid behavior have 
been found in some organic one-dimensional conductors \cite{3} and in the normal 
state of high temperature superconductors \cite{4}. A whole variety of 
experimental techniques have been employed in the search for such behavior, 
including resistivity measurements \cite{5}, infra-red spectroscopy \cite{6}, 
scanning tunneling spectroscopy \cite{7} and time-resolved two-photon 
photoemission \cite{8}. Angle-resolved photoemission (ARPES) has an advantage, 
in that the energy and lifetime of the photo-hole are directly observable in the 
experiment. ARPES in principle measures the quasiparticle spectral function 
\cite{9}:
\begin{equation}
\mathrm{A}(\bf{k},\omega)\propto \frac{{\mathrm Im}\Sigma({\bf k},\omega)}{[\omega-\epsilon_{{\bf k}}-\mathrm{Re}\Sigma({\bf k},\omega)]^{2}+({\mathrm Im}\Sigma({\bf k},\omega))^{2}}
\label{eq:1}
\end{equation}
where $\epsilon_{\bf{k}}$ represents the energy of the state in the Hartree potential and $\Sigma({\bf k},\omega)$
is the quasiparticle self-energy reflecting many body interactions. Thus, 
momentum resolved self-energies are directly accessible in the experiment and as 
such, ARPES represents a crucial experimental probe for the presence or absence 
of Fermi liquid behavior. Furthermore, complications connected to the lifetime 
of the photoelectron (in three-dimensional systems) may be overcome in quasi 
low-dimensional systems. Indeed, there have already been several photoemission 
studies, which quantitatively compare peak widths to calculated lifetimes for 
metallic surface states \cite{10} and two-dimensional states in layered 
materials \cite{11}.

When considering the lifetime of the valence hole, there are three main decay 
mechanisms: electron-electron scattering, electron-phonon scattering and 
impurity (defect) scattering. In three-dimensional systems, the {\it electron-
electron} interaction contributes to the total width or inverse lifetime with the 
term $\Gamma_{e-e}(\omega,T)=2\beta[(\pi k_{B}T)^{2}+\omega^{2}]$. This scattering rate does not depend on the form of the interaction, but 
it may depend on the shape of the Fermi surface. If the scattering process is 
two-dimensional, then the quadratic energy (temperature) dependence is modified 
by a logarithmic factor \cite{12}. Previous attempts to observe this term for 
"prototypical" Fermi liquids, free-electron-like metals, have failed because $\beta$ is often too small. Indeed, with estimated values of $~10^{-2}  \mathrm{eV}^{-1}$ for $\beta$, $\Gamma_{e-e}$ 
contributes less than 5 meV variation through the whole band for the surface 
states studied \cite{10}. Further, the temperature dependent contribution is 
negligible, being of the order of 0.1 meV at room temperature. 
The {\it electron-phonon} scattering contribution to the inverse lifetime
$\Gamma_{e-p}(\omega,{\mathrm T})$ is given by: 
\begin{equation}
2\pi \int_{0}^{\infty}d\nu \alpha^2F(\nu)[2n(\nu)+f(\nu+\omega)+f(\nu -\omega)]
\label{eq2}
\end{equation}    
where $\alpha^{2}F(\omega)$ is the Eliashberg coupling function, and $f(\omega)$ and $n(\omega)$ are the 
Fermi and Bose-Einstein functions. This term monotonically increases with energy 
over the region $|\omega|<\omega_{max}$ (for T=0), where $\omega_{max}$ is the cutoff of the phonon 
spectrum. The exact functional form is slightly dependent on the phonon 
spectrum. The temperature dependence of $\Gamma_{e-p}$ is approximately linear at higher 
temperatures, with slope $2\pi\lambda k_{B}$, where $\lambda$ is the electron-phonon coupling 
constant. For most metals 20 meV$<\omega<$100 meV, and $\lambda$ falls between 0.1 and 1.5. 
{\it Impurity} scattering is elastic provided that the energy gap between the impurity 
ground state and its lowest excited state is large compared with the hole 
energy. To a first approximation, it is proportional to the impurity 
concentration, but independent of energy or temperature. Thus, at sufficiently 
low temperature in any real system, it will be the dominant decay mechanism for 
the hole close to $E_{F}$. 
If the scattering mechanisms are independent, the total scattering rate is given 
by $\Gamma_{tot} = \Gamma_{e-e} + \Gamma_{e-p} + \Gamma_{e-i}$. Virtually all temperature dependence in $\Gamma_{tot}$ reflects the $\Gamma_{e-p}$ term, whilst any energy dependence falls within two distinct 
regions. In a narrow region $|\omega|\leq\omega_{D}$ (at T=0) the only significantly varying 
term is $\Gamma_{e-p}$. Any measurable change out of that region, is most likely due to 
the $\Gamma_{e-e}$ term. The real part of the quasiparticle self-energy is also known for 
all three mechanisms of scattering. Of these, electron-phonon coupling is the 
only term which is able to alter the quasiparticle mass in a narrow energy 
region around the Fermi level. 

In the present paper, we report a detailed study of the lifetime effects 
associated with photo-excitation of a surface state on the Mo(110) surface. The 
electronic structure of this surface has already been the subject of several 
studies \cite{13}. Our own tight-binding calculations in the slab formulation 
reproduce the presence of $d$-derived surface states or resonances in the mid-
region of the $\overline{\Gamma}-\overline{\mathrm{N}}$ and $\overline{\Gamma}-\overline{\mathrm{H}}$ lines, and indicate a bandwidth of the order of 1-2 eV. It 
is well established that the line-shape or width of a photoemission peak 
associated with excitation from such states reflects the lifetime of the photo-
hole and is independent of any momentum broadening associated with the 
photoelectron \cite{14}. However, the fact that bulk states are available for 
the surface state photo-hole to scatter into, means that the three-dimensional 
decay rates are applicable even in the present case.

The experiment was carried out at the National Synchrotron Light Source, using 
the VUV undulator beamline U13UB, which is based on a 3-m normal incidence 
monochromator. The photon energy used in the study was 15.16 eV. The electron 
analyzer was a Scienta SES-200, which uses a two dimensional micro-channel plate 
as a detector. The detector collects simultaneously a wide energy window and a 
wide angular window ($\sim12^{\circ}$) of excited photoelectrons. This greatly reduces the 
time needed for data acquisition. The spread of the UV light was less than 2 
meV, and the energy resolution of the electron analyzer was around 5 meV. The 
combined energy resolution therefore makes a minimal contribution to widths 
observed in this study. The angular resolution is of the order of $0.1^{\circ}$ which 
contributes 10 to 15 meV to the width of the studied state. The sample was cleaned by oxidation cycles (at 1400 K in an 
oxygen atmosphere of $~5\times 10^{-6}$ Pa), followed by flashes to $~2200$ K. The sample 
temperature was measured with a W-WRe thermocouple. The base pressure was $4\times 10^{-9}$ Pa, more than 90 \% of which was hydrogen.

\begin{figure}
\centerline{\epsfxsize=9.4cm\epsfbox{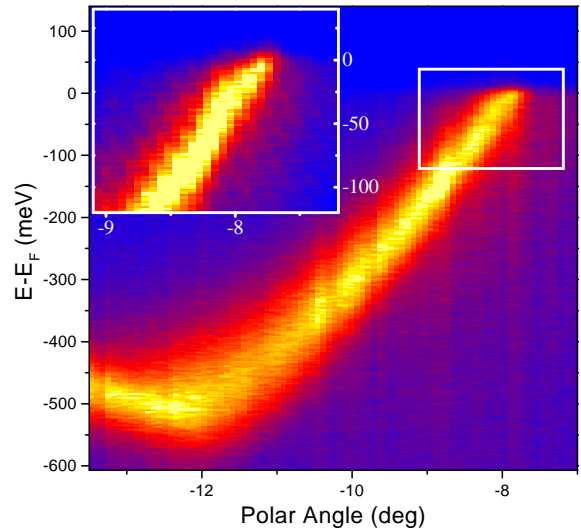}}
\caption{
ARPES intensity plot of the Mo(110) surface recorded along the $\overline{\Gamma}-\overline{\mathrm{N}}$ line of the SBZ 
at 70 K Shown in the inset is the spectrum of the region around $k_{F}$ taken with 
special attention to the surface cleanliness.
}
\label{fig:1}
\end{figure}

Fig. \ref{fig:1} shows a typical spectrum of the Mo(110) surface, taken along the $\overline{\Gamma}-\overline{\mathrm{N}}$ symmetry line, with the sample held at 70 K. The state shown in the figure corresponds to the surface resonance which closes the elliptical hole Fermi 
orbit around $\overline{\Gamma}$ \cite{13}. The spectrum in the inset that shows a narrow energy 
region close to the Fermi level was taken in a short time interval, 1 min. Such 
short measuring intervals are required because, as shown later, there is a 
significant influence from adsorption of residual gasses. Important qualitative 
observations can be made directly from the spectrum shown in the inset: the 
state sharpens up on going towards the Fermi level, and there is an obvious 
change in band velocity near $E_{F}$. We focus our attention on these observations. 
Quantitative analysis is performed by taking slices through the spectrum, either 
at constant emission angle, or at constant energy to obtain the spectral 
intensity as a function of energy or emission angle, respectively. The two 
methods are equivalent in the sense that they provide information on the same 
A$(\bf{k},\omega)$. However, analyzing the data by taking slices at constant energies is 
often more convenient because it is not affected by the Fermi distribution in 
the same way that slices at constant angle are. In our analysis we have used 
"horizontal" cuts (at constant energy) to extract the dispersion, and "vertical" 
cuts (at constant angle) to extract the width of the quasiparticle peak. 

\begin{figure}
\centerline{\epsfxsize=7cm\epsfbox{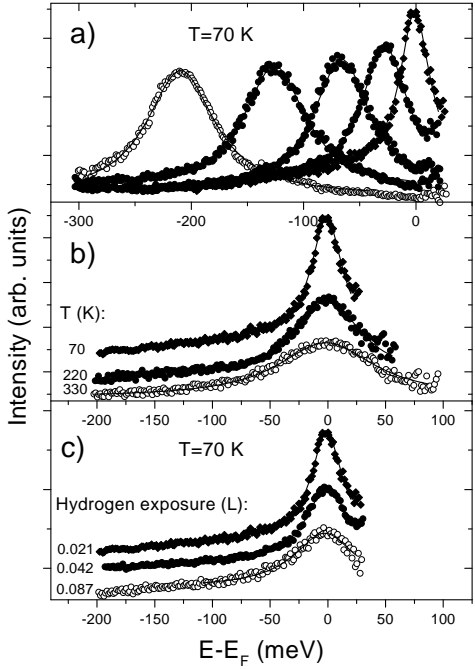}}
\caption{
Spectral intensity as a function of binding energy for constant emission angle, 
normalized to the experimentally determined Fermi cut-off. Data are symbols, 
while lines are fits to the Lorentzian peaks with a linear background. The 
dependence on the binding energy (a), temperature (b), and hydrogen exposure (c) 
is shown.
}
\label{fig:2}
\end{figure}

Fig. \ref{fig:2} shows representative spectra obtained by slicing the spectral intensity 
in the "vertical" direction. Spectra are divided by an experimentally determined 
Fermi cut-off. The analysis was performed by fitting the spectra to Lorentzian 
peaks with a small linear background. 
The high-energy limit for the fitting 
interval was always below $3k_{B}T$. Influences of the three experimental 
parameters; binding energy (a), temperature (b) and hydrogen exposure (c) are 
shown, one being varied, as the others two are kept constant. The trends are 
obvious; the width increases with all three parameters.

\begin{figure}
\centerline{\epsfxsize=9cm\epsfbox{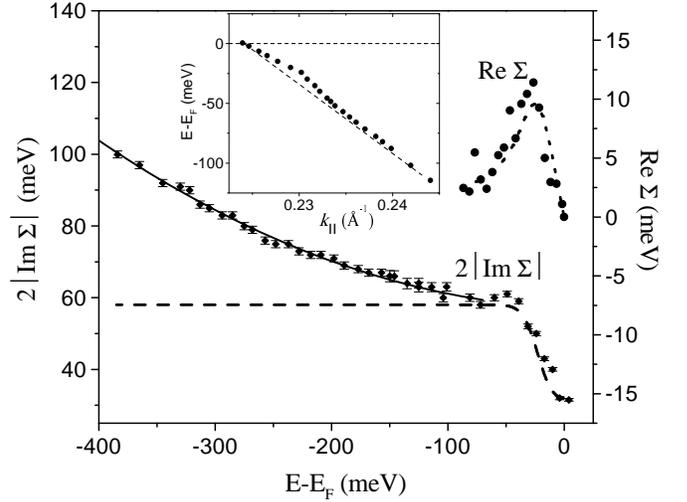}}
\caption{
The photo-hole self-energy as a function of binding energy at 70 K. The real 
part is obtained from the dispersion. The imaginary part is obtained from the 
width of the quasiparticle peak. The solid line is a quadratic fit to the high 
binding energy data ($\omega<-80$ meV). The dashed (dotted) line shows the calculated 
electron-phonon contribution to the imaginary (real) part of the self-energy. 
The dashed line is shifted up by 26 meV.
}
\label{fig:3}
\end{figure}

The energy dependence of the quasiparticle peak width is shown in Fig. \ref{fig:3}. The 
error bars are statistical uncertainties from the fits to peaks such as those 
shown in Fig. \ref{fig:2}(a). The peak width shows a minimum at $\omega=0$, a sharp increase 
within the interval $0>\omega>-40$ meV, followed by a slower increase at higher binding 
energies. The same behavior is also found for the angular peak width 
("horizontal" cuts). At the same temperature, all spectra show the same 
dependence, offset by a constant value dependent on the surface impurity level. 
The sharp increase in width near the Fermi level reflects the electron-phonon 
scattering. This is confirmed if we compare the experimental points with $\Gamma_{e-p}$ 
calculated from equation (2), using a theoretical $\alpha^{2}F(\omega)$ for bulk molybdenum 
\cite{15}. Both the range and magnitude of the increase agree well with 
calculation, suggesting that the theoretical bulk electron-phonon coupling 
constant from ref. \cite{15}, $\lambda=0.42$, applies also to the surface studied 
here. Further, the surface Debye energy is similar to the bulk value ($\approx 30$ meV), 
in accordance with recent measurements of surface phonon dispersions \cite{16}. 
To obtain the agreement between the experimental points and the theoretical 
curve it was necessary to shift the latter uniformly by 26 meV. We attribute 
this difference to impurity scattering. The observation that the experimental 
points shift uniformly by the same amount is an indication that the impurity 
scattering is independent of energy. If we subtract the calculated electron-
phonon contribution from the total width, we are left with a monotonic increase 
in binding energy that can be fitted with a parabola. This component is the 
electron-electron scattering term, with coefficient $2\beta\approx 0.28 \mathrm{ eV}^{-1}$. Within the Born 
approximation, $2\beta\approx (\pi U^{2})/(2W^{3})$. Our result is consistent with the on-site Coulomb repulsion 
$U\approx 0.6$ eV predicted for molybdenum \cite{17} if we equate $W$ with the bandwidth of 
the surface state ($\approx 1.3$ eV). It should be noted that for such a rapidly 
dispersing band with negative velocity, the measured energy width could be 
smaller than the intrinsic width \cite{18}. We estimate that in our case this 
"compression" of the spectral width is less than 7 \%, which has minor 
consequences on our results. 

Also shown in Fig. \ref{fig:3} is the real part of the quasiparticle self-energy. 
Experimentally, it is obtained by subtracting a straight line from the measured 
dispersion of the quasiparticle peak (as shown in the inset of Fig. \ref{fig:3}). In this 
energy range the straight line represents a good approximation for the 
dispersion of the "non-interacting" system (i.e. the system without electron-
phonon coupling). It is chosen to have the same $k_{F}$ as the quasiparticle 
dispersion (Luttinger theorem for interacting fermions) and to match the 
quasiparticle dispersion in the range $\omega\approx -100$ meV. This procedure gives only the electron-
phonon term for the real part of the self-energy. The component reflecting the 
electron-electron interaction stays hidden in the dispersion of the "non-
interacting" system. Also shown is the electron-phonon contribution to the real 
part of the self-energy, obtained via the Kramers-Kronig transformation of the 
calculated imaginary part.

\begin{figure}
\centerline{\epsfxsize=8.5cm\epsfbox{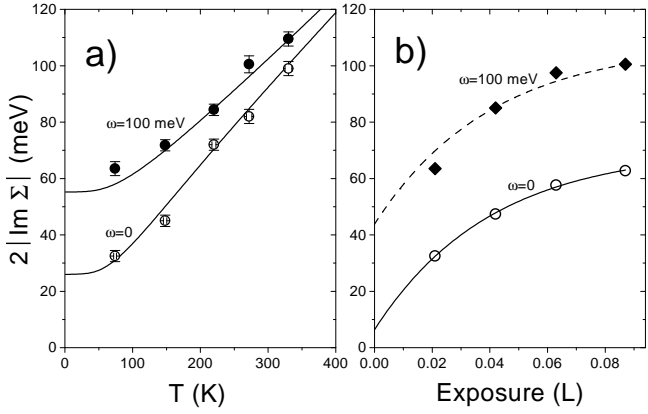}}
\caption{
The peak width as a function of temperature (a) and hydrogen exposure (b), 
measured for two binding energies. For the exposure dependence the sample was 
held at 70 K. Lines in a) are calculated electron-phonon contributions, shifted 
up by 26 meV to match the data. Lines in b) represent fits (see text for 
details).
}
\label{fig:4}
\end{figure}

Fig. \ref{fig:4} shows the width of the quasiparticle peak as a function of temperature 
(a) and as a function of hydrogen exposure $\theta$ (b) for two different binding 
energies. In Fig. \ref{fig:4}(a), we also show electron-phonon contributions calculated 
from (2), and shifted up by 26 meV. The excellent agreement with the 
experimental points confirms that the temperature dependence reflects the 
electron-phonon interaction. Linear fits to the experimental data points produce 
different values for the electron-phonon coupling constant for the two binding 
energies: $\lambda=0.52$ (at $\omega=0$) and $\lambda=0.35$ (at $\omega=100$ meV), compared to the theoretical 
value of 0.42.

The observation that the width of the quasiparticle peak always has a 
significant constant term indicating the presence of impurity scattering lead to 
further investigation. It is known that this surface state is very sensitive to 
hydrogen adsorption. Fig. \ref{fig:4}(b) shows how the width changes with 
the exposure to residual hydrogen. Note that it saturates with exposure 
$\theta$. If the scattering rate is proportional to the concentration of adsorbed 
particles, the experimental points become a measure of the concentration. Since 
the number of free adsorption sites decays exponentially with exposure, the 
concentration of adsorbed atoms as a function of exposure should change as 
$c(\theta)=c_{0}+c_{sat}(1-e^{-p\theta})$, where $p$ is the adsorption probability and $c_{0}$ ($c_{sat}$) is the 
initial (saturation) concentration. The width of the quasiparticle peak can be 
fitted with the same dependence (lines). It is notable that extrapolation 
to zero exposure results in a residual width of $6\pm 5$ meV at $\omega=0$. Electron-phonon 
coupling contributes with $\approx 5$ meV for T=70 K. However, we should also note that 
there is some uncertainty in the initial coverage due to the change in 
adsorption conditions between flashing the sample and the measurement.

In conclusion, we have analyzed the dispersion and the width of the Mo(110) 
surface state and isolated different mechanisms for scattering of the 
quasiparticle. For the first time it has been possible to isolate the electron-
electron, electron-phonon and electron-impurity scattering contributions to the 
quasiparticle lifetime. The electron-electron contribution is shown to be an 
order of magnitude higher than that for $s-p$-derived states. Our study shows that 
ARPES offers the possibility of momentum resolving the electron-phonon 
contribution to the real and imaginary parts of the self-energy.

This work is supported by the U.S. Department of Energy (DOE) under Contract No. 
DE-AC02-98CH10886.


\begin{references}
\bibitem{1} D. Pines and P. Nozi\`{e}res, {\it The Theory of Quantum Liquids} (Benjamin, 
New York, 1969).

\bibitem{2} G. Grimvall, {\it The Electron-Phonon Interaction in Metals} (North-
Holland, New York, 1981).

\bibitem{3} C. Bourbonnais {\it et al}, J. Phys. Lett. {\bf 45} L-755 (1984); B. 
Dardel {\it et al}, Europhys. Lett. {\bf 24}, 687 (1993).

\bibitem{4} C. G. Olson {\it et al}, Phys. Rev. B {\bf 42}, 381 (1990).

\bibitem{5} M. Gurvitch and A. T. Fiory, Phys. Rev. Lett. {\bf 59}, 1337 (1987).

\bibitem{6} G. A. Tomas {\it et al}, Phys. Rev. Lett. {\bf 61}, 1313 (1988).

\bibitem{7} J. Li, {\it et al}, Phys. Rev. Lett. {\bf 81}, 4464 (1998).

\bibitem{8} W. Nessler {\it et al}, Phys. Rev. Lett. {\bf 81}, 4480 (1998).

\bibitem{9} In sudden approximation ARPES intensity is given by 
I$({\bf k},\omega)=|{\mathrm M}({\bf k},\omega)|^{2}{\mathrm A}({\bf k},\omega)f(\omega)$, where A$({\bf k},\omega)$ is the spectral function, $f(\omega)$ is 
the Fermi function and M$({\bf k},\omega)$ is a slowly varying term containing matrix 
elements for photoemission.

\bibitem{10} B. A. McDougall, T. Balasubramanian and E. Jensen, Phys. Rev. B 
{\bf 51}, 13891 (1995); T. Balasubramanian {\it et al}, Phys. Rev B {\bf 57}, R6866 
(1998); P. Hofmann {\it et al}, Phys. Rev. Lett. {\bf 81}, 1670 (1998).

\bibitem{11} R. Claessen {\it et al}, Phys. Rev. Lett. {\bf 69}, 808 (1992).

\bibitem{12} C. Hodges, H. Smith and J. W. Wilkins, Phys. Rev. B {\bf 4}, 302 
(1971).

\bibitem{13} K. Jeong, R H. Gaylord and S. D. Kevan, Phys. Rev. B {\bf 38}, 
10302 (1988); K. Jeong, R H. Gaylord and S. D. Kevan, Phys. Rev. B {\bf 39}, 
2973 (1989).

\bibitem{14} J. B. Pendry in {\it Photoemission and the Electronic Properties of 
Surfaces}, B. Feuerbacher, B. Fitton and R. F. Willis (Wiley, New York, 1978).

\bibitem{15} S. Y. Savrasov and D. Y. Savrasov, Phys. Rev. B {\bf 54}, 16487 
(1996).

\bibitem{16} J. Kr\"{o}ger, S. Lehwald and H. Ibach, Phys. Rev. B {\bf 55}, 10895 
(1997).

\bibitem{17} W. A. Harrison, {\it Electronic Structure and the Properties of Solids} 
(W. H. Freeman \& Co, San Francisco, 1980).

\bibitem{18} E. D. Hansen, T. Miller and T. -C Chiang, Phys. Rev. Lett. {\bf 
80}, 1766 (1998).
\end{references}
\end{document}